# A Comparative Study of Straight-Strip and Zigzag-Interleaved Anode Patterns for MPGD Readouts

C.Perez-Lara[2], S. Aune[3], B. Azmoun[1], K. Dehmelt[2], A. Deshpande[2], W. Fan[2], P. Garg[2], T. K. Hemmick[2], M. Kebbiri[3], A. Kiselev[1], I. Mandjavidze[3], M. L. Purschke[1], M. Revolle[4], M. Vandenbroucke[4], C. Woody[1]

*Abstract*—Due to their simplicity and versatility of design, straight strip or rectangular pad anode structures are frequently employed with micro-pattern gas detectors to reconstruct high precision space points for various tracking applications. The particle impact point is typically determined by interpolating the charge collected by several neighboring pads. However, to effectively extract the inherent positional information, the lateral spacing of the straight pads must be significantly smaller than the extent of the charge cloud. In contrast, highly interleaved anode patterns, such as zigzags, can adequately sample the charge with a pitch comparable to the size of the charge cloud or even larger. This has the considerable advantage of providing the same performance while requiring far fewer instrumented channels. Additionally, the geometric parameters defining such zigzag structures may be tuned to provide a uniform detector response without the need for so-called "pad response functions", while simultaneously maintaining excellent position resolution. We have measured the position resolution of a variety of zigzag shaped anode patterns optimized for various MPGDs, including GEM, Micromegas, and μRWELL and compared this performance to the same detectors equipped with straight pads of varying pitch. We report on the performance results of each readout structure, evaluated under identical conditions in a test beam.

## I. INTRODUCTION

The rising popularity of micropattern gas detectors (MPGDs) over the last two decades in high energy and nuclear physics experiments is due in large part to their ability to deliver excellent position resolution, well below 100μm, while allowing for great flexibility in design [1 - 3]. To achieve this level of performance, the readout plane often is segmented into many straight strips or rectangular pads [4] with effectively a single operant parameter, the pitch. To achieve a given position resolution specification, the pitch must simply be made small enough such that several contiguous pads adequately sample the incident charge. Unfortunately, this usually requires relatively large pad densities and in turn high cost related to the number of instrumented readout channels. As an alternative, highly interleaved zigzag shaped charge collecting anodes sufficiently sample charge clouds roughly the size of the pitch or even smaller, resulting in a highly efficient position encoding mechanism [5 – 17]. Unlike straight strips, multiple operant parameters govern the division of charge among neighboring zigzag pads, enabling the readout to be precisely tailored for specific detector configurations. GEM detectors equipped with a zigzag readout have already been shown to have equal or superior performance compared to straight pads with almost five times smaller pitch and no reliance on pad response functions [5]. As a result, zigzag shaped anodes would be ideal candidates for use in future tracking systems, such as at the future electron ion collider (EIC) at Brookhaven National Laboratory [18], which is the long term goal of this study.

We have evaluated multiple MPGD detectors equipped with many variants of a zigzag readout design, each distinguished by specific combinations of the pitch, zigzag period and the degree of interleaving, otherwise referred to as the "stretch" parameter, defined by the degree of overlapping between neighboring pads (as a percent of the pitch). This allowed for a relatively comprehensive scan of the relevant region of the zigzag parameter space, leading to the identification of optimal parameter sets with uniform response and excellent position resolution [5, 7]. The same detectors were also read out with several straight pad anodes of varying pitch, under virtually identical conditions for the purpose of comparison.

## II. EXPERIMENTAL SETUP

A beam test was conducted at the Fermilab test beam facility (FTBF) [19] in March 2019 to characterize the performance of a multitude of zigzag anode patterns used to read out quadruple-GEM, Micromegas, and μRWELL detectors in a planar tracker configuration. The readout PCB for each detector comprised an array of 100 1 cm × 1 cm regions (or "cells"), each containing 3 to 15 zigzag pads with a unique geometry. As shown in Fig. 1, ten cells were reserved for straight pad patterns of varying pitch, allowing for a direct performance comparison, under identical conditions using a single board. The pitch of the zigzags and straight pads varied from about 0.4mm – 3.33mm, the zigzag period ranged from 0.33mm – 1.0mm, and the stretch parameter spanned a range from -25% - 25% for all the boards. In all cases, the pads were 10mm long to completely fill each cell.

Every cell of each chamber was exposed to a normally incident beam of 120GeV protons, which produced about 149 primary ionization electrons in a 12.5 mm drift gap in Ar:CO (30%) and 157 electrons in Ar:iC$_4$H$_{10}$ (5%), in the case of the GEM, and

Manuscript submitted on May 7, 2020. This work was supported in part by the U.S. Department of Energy under Prime Contract No. DE-SC0012704.

[1] Brookhaven National Laboratory, Physics Department, Upton, New York, United States of America

[2] Stony Brook University, Department of Physics and Astronomy, Stony Brook, New York, United States of America
[3] CEA Saclay, DRF/IRFU/DEDIP, Gif-sur-Yvette, France
[4] CEA Saclay, DRF/IRFU/DPHN, Gif-sur-Yvette, France

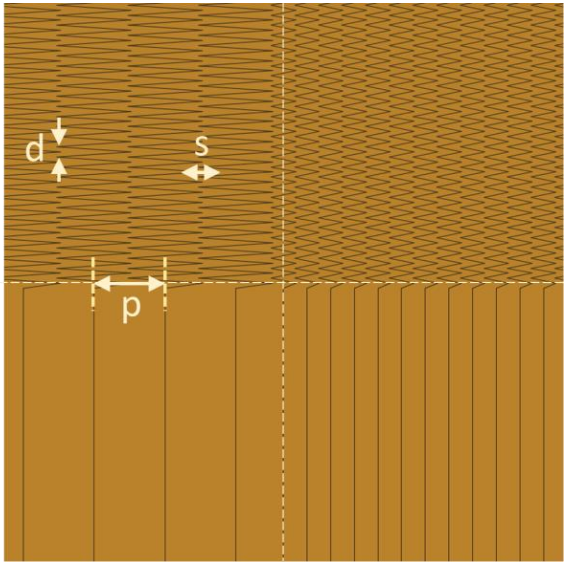

**Fig. 1** Zoom-in on PCB pad plane, depicting a pair of straight and zigzag anode patterns of varying pitch, denoted by "p". In the case of the zigzags, "d" denotes the zigzag period, and "s" is the stretch parameter, which shows the degree of the interleaving (measured as the overlapping area of adjacent triangles). The anode pads were formed using a laser ablation process, resulting in an inter-pad spacing of about 20μm.

Micromegas and μRWELL detectors, respectively. Charge collected by each anode was amplified and digitized using DREAM front end electronics [20, 21], operating at a sampling rate of 16.6MHz with sixteen time samples written to disk. The gain was set appropriately high to fully utilize the dynamic range and to maintain a noise to signal ratio of about 2% on each readout channel.

A charge weighted average of the fired pads (or centroid) was computed to measure the particle impact point. A high resolution silicon telescope [22, 23] just upstream of this setup was also used to measure reference tracks to a precision of about 20 μm for the purpose of determining the position resolution. A sketch of the experimental apparatus is shown in Fig.2 and Table 1 summarizes the experimental parameters used for this setup. (Note: the dissimilar gains listed in this table are expected to be inconsequential since the position resolution does not improve beyond a gain of about $10^4$ for similar detectors tested in [5].)

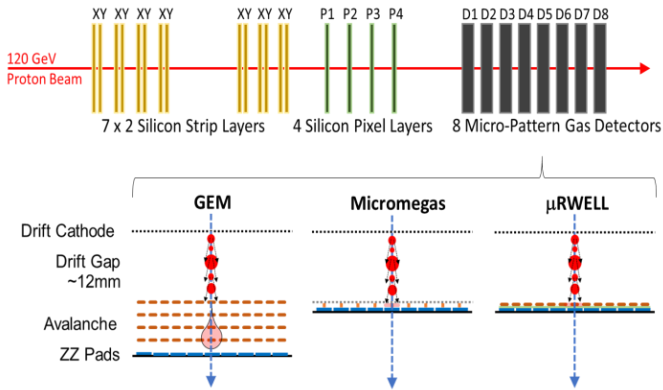

**Fig. 2** Top: Layout of experimental apparatus at FTBF. Bottom: Detector configuration of each kind of MPGD detector.

**Table 1 MPGD operating parameters**

| MPGD | Gas | Gain | Gain Element Potential | Drift Gap | Transfer Gaps |
|---|---|---|---|---|---|
| 4-GEM (Std. GEMs) | Ar+CO$_2$ (30%) | 2x10$^4$ | 350V | 12.3mm 1.2kV/cm | 1.6mm 3kV/cm |
| Micromegas (Non-resistive) | Ar+IsoBut (5%) | 1.5x10$^4$ | 700V | 12.3mm 1.2kV/cm | NA |
| Std. μRWELL | Ar+IsoBut (5%) | 1x10$^4$ | 560V | 12.3mm 1.2kV/cm | NA |

### III. RESULTS

#### A. Position Correlation

The set of plots in Fig. 3 provide a direct performance comparison between straight and zigzag pads (for the case of a 4-GEM detector) for a pad pitch ranging from 0.4mm to 2mm. Once the coordinate systems of the GEM detector and the Si telescope are aligned using an offline procedure, as described in [5], each plot correlates the calculated centroid and the beam position, as measured with the telescope. The results for the zigzag pattern exhibiting optimal performance in terms of resolution and linearity of response is shown.

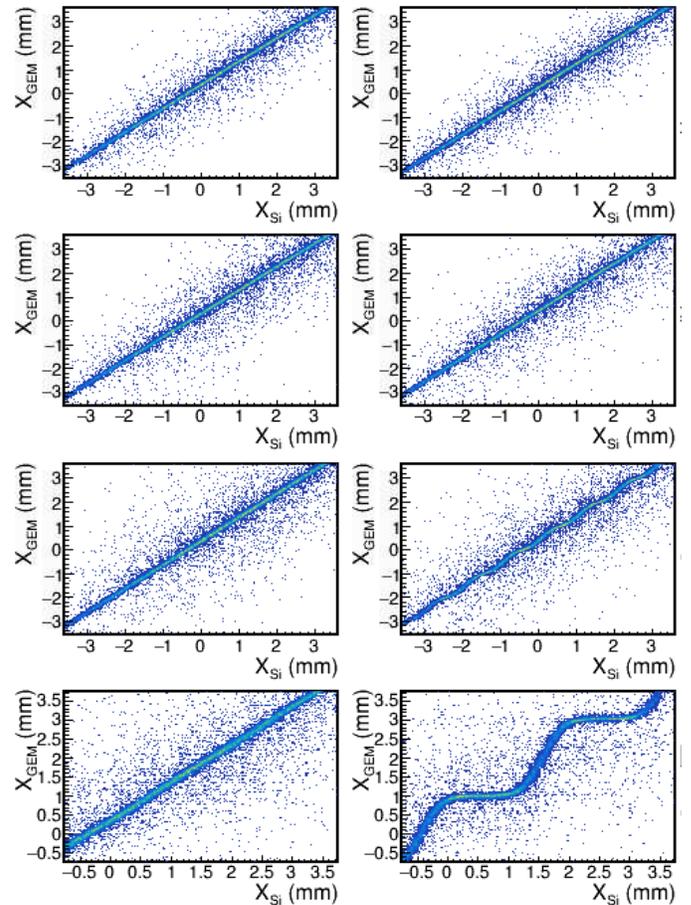

**Fig. 3** The calculated centroid vs the beam position (measured with the silicon telescope) for zigzag (left) and straight (right) pads, for the case of 0.4 mm, 0.67 mm, 1 mm, and 2 mm pitch, respectively from top to bottom.

For the straight strips, these plots reveal an obvious oscillation and prominent plateau regions for the case of 1mm and 2mm pitch, respectively. Both plots indicate a significant non-linear response of the readout while the latter indicates areas with no positional sensitivity at all. In the plateau regions, the resolution is effectively equal to the linear extent divided by $\sqrt{12}$. In contrast, the zigzag pads demonstrate particularly good linear and uniform response across multiple pitch cycles for all the pitch values studied. Not until the pitch is reduced to well below 1mm, do the results from the straight pad anode reveal a linear response.

The flat, insensitive regions of the straight pad response correspond to the center of the pad, where charge sharing tends to be poor. Once the lateral extent of a straight pad is comparable in size to the incident charge cloud, the vast majority of the charge is collected onto a single readout channel, making interpolation impossible. This is evident in Fig.4, which shows the relative degree of charge sharing as a function of different positions of the pad. In this case, the maximum pad charge to the total cluster charge ratio is flat over the exact regions corresponding to the insensitive, plateau regions in Fig. 3. The same ratio for the case of the zigzag pads is substantially smaller at larger pitch, an indication of superior charge sharing, where no distinct plateau is ever observed. The reason this ratio does not top out at unity in the case of straight pads is believed to be partly due to inter-pad cross-talk, which symmetrically induces a bipolar signal on neighboring pads. However, the effective contribution from this induced signal was measured to be only at the 2-3 % level. It is believed the collection of small amounts of peripheral charge due to a low zero suppression threshold is responsible for the remaining charge deficit. Ultimately, this effect was captured by all the readout patterns tested and the measured resolution appears to be minimally impacted by this.

### B. Position Resolution

The resolution versus various anode parameters for both zigzag and straight pads is shown in Fig. 5, for all three MPGDs under investigation. As indicated in the figure, the quoted resolution for the zigzag pads in each plot is derived from a Gaussian fit to the raw residual distribution, an example of which is shown in Fig. 6. However, for the straight strips, since the response is highly non-uniform and pitch dependent, it was necessary to decompose the residual distribution into single-hit and multi-hit components, also shown in Fig 6, in order to make a useful comparison. Furthermore, in the case of multi-pad hits, a pad response function was employed to linearize the straight pad response to improve the resolution. The effective resolution for the straight pads was then taken to be the weighted average of the RMS of the single-pad hit distribution and the sigma of the corrected multi-pad hit distribution, fitted to a simple Gaussian. The straight strip component resolutions are also plotted in Fig. 5 for comparison. In the case of the zigzags, such a correction procedure is unnecessary since the response is already linear.

Though the resolution for both anode patterns converges at lower pitch, it is clear the zigzags maintain high performance at significantly larger pitch. As mentioned earlier, the single and multi-hit components of the straight pad residual distributions are spatially separated, where single pad hits correspond to collection sites near the pad center and multi-hits to sites in between pads. Near the midline region between pads, the straight pad resolution is similar to that of the zigzags, however the resolution quickly degrades to the point of total insensitivity at the very center of a pad. For a 2mm pitch, single pad hits span an area a little more than half the pitch, with an expected resolution of ~ 1 mm / $\sqrt{12}$, which is in line with the measured single-pad hit RMS of about 300 μm shown in Fig. 6.

The comparatively worse resolution observed for Micromegas is likely related to spatial distortions induced by the pillars on the collected charge. This effect seems to be amplified in regions very close to the pillars, whereas the resolution was found to be on par with GEMs and μRWELL in small regions far away from the pillars.

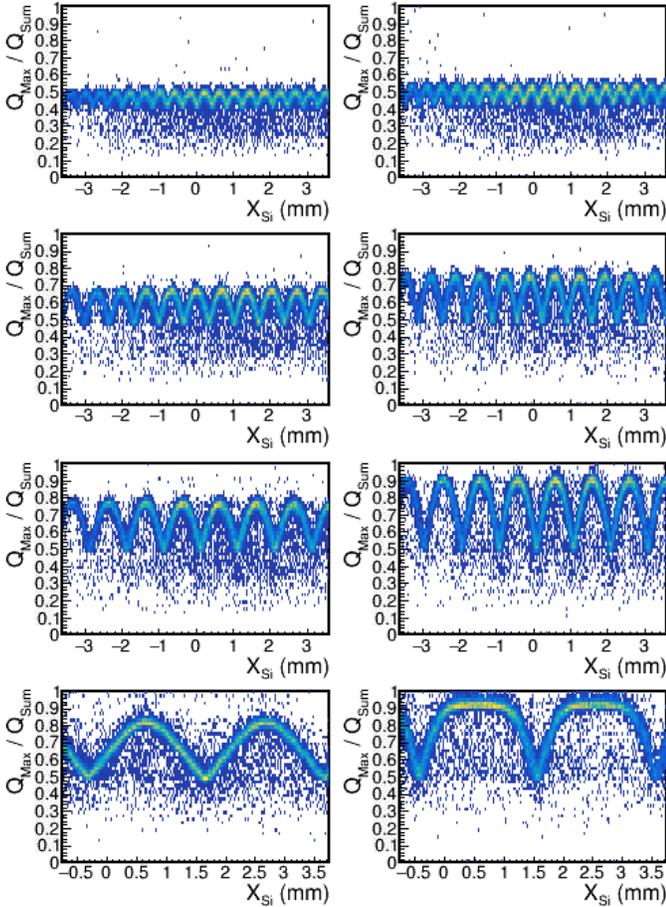

**Fig. 4 Ratio of the maximum pad charge to the total cluster charge for both zigzag (left) and straight pads (right), for the case of 0.4 mm, 0.67 mm, 1 mm, and 2 mm pitch, respectively from top to bottom.**

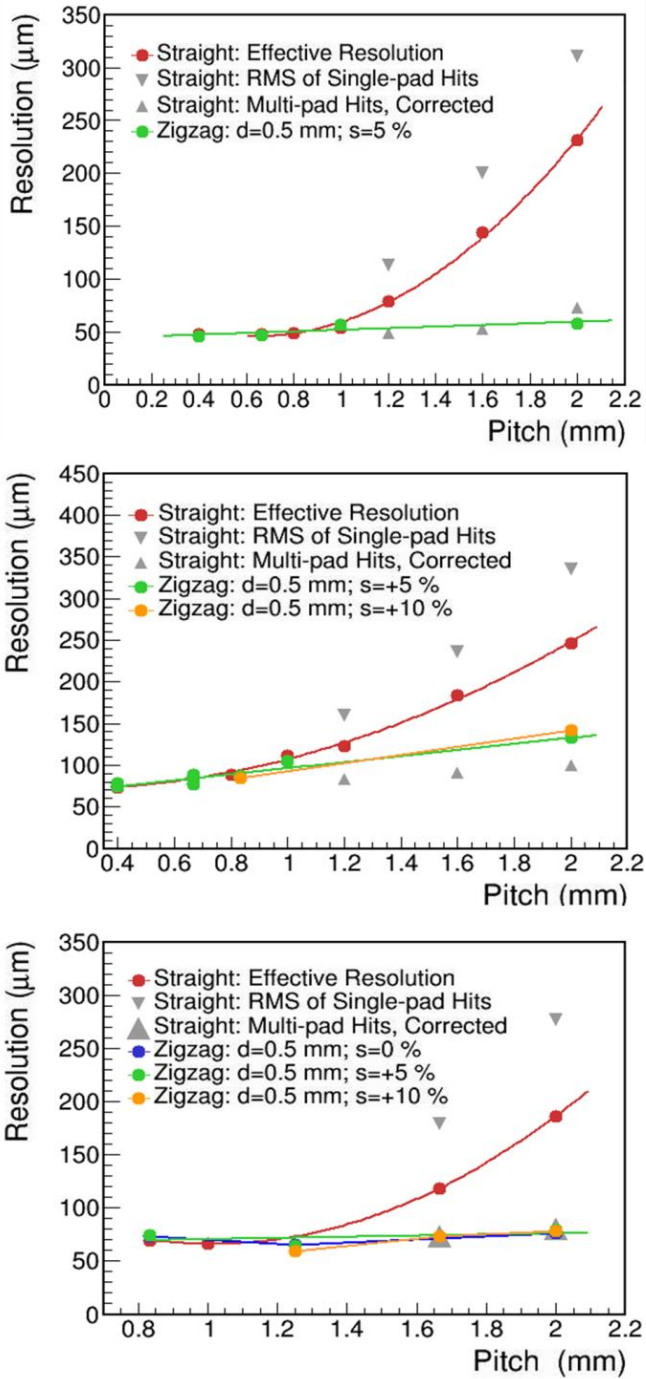

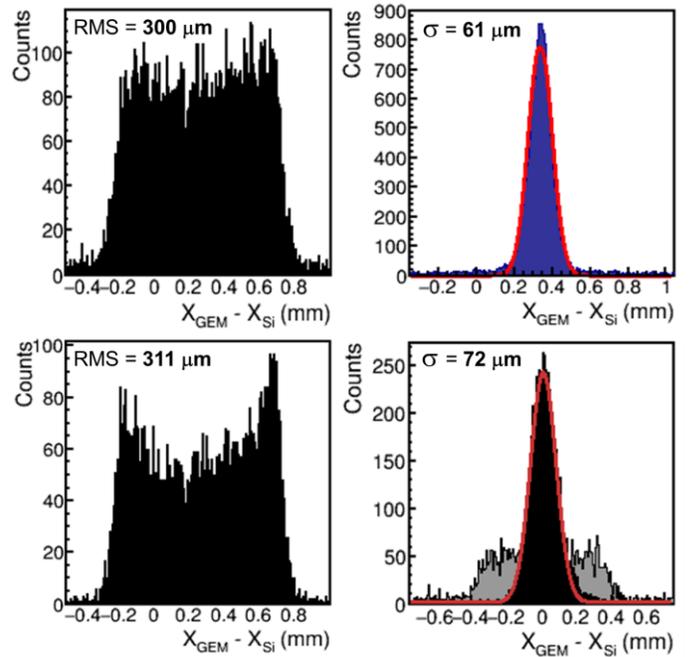

**Fig. 6** The black histograms correspond to a straight pad readout with 2mm pitch. The upper left panel shows the raw residual distribution, which is decomposed into single and multi-pad residual distributions, shown respectively in the bottom two panels. The multi-hit residuals were corrected and fit to a Gaussian, shown in red and the gray histogram shows the pre-correction residual spread. The blue histogram corresponds to the raw residual distribution from a zigzag pattern with 2mm pitch, fit to Gaussian.

*C. Pad Merging*

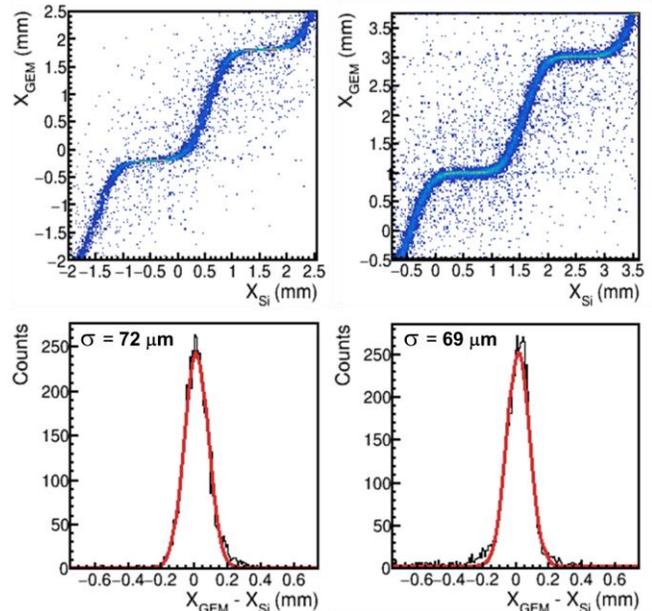

**Fig. 5** Position resolution vs pitch for both zigzag and straight pad readouts of a 4GEM, Micromegas, and µRWELL detector, respectively. In the case of the zigzags, the resolution is measured for zigzag patterns with various period and stretch parameters.

For most applications, trails of ionization left behind by particle tracks are not necessarily localized over small regions of the readout, as in the case of normally incident tracks, therefore, this phenomenon, in general does not pose a major concern.

**Fig. 7** The left two plots correspond to data taken with straight pads with 0.4 mm pitch. Charge from blocks of 5 contiguous pads was merged to recreate the response of straight pads with 2mm pitch. The results on the right correspond to a direct measurement with straight strips with 2mm pitch. (The residual distributions exclude single pad hits.)

It must be noted that a pad merging method was adopted to generate two data points in each of the plots in Fig. 5: at 1.2 and 1.6 mm for the GEM and Micromegas, and at 1.6 and 2.0 mm for the µRWELL detectors. Charge collected by finely pitched straight strips was effectively re-binned, event by event, by factors of two, three, and four in order to recreate the corresponding charge distribution resulting from coarse segmentations of the readout. In Fig. 7 this scheme is applied to straight strips with a pitch of 0.4 mm to recreate the response of a 2mm pitch readout. Upon comparing these results to a direct measurement with a 2mm pitch readout, the agreement is remarkably good and confirms the validity of this exercise.

## IV. Summary

GEM detectors read out with appropriately formulated zigzag shaped anodes, including relatively coarse segmentations, have been shown to have excellent position resolution and a virtually uniform detector response with no need for pad response corrections. As a result, the channel count may be lowered substantially compared to detectors using more conventional straight pad readouts, which typically require much higher channel densities to maintain the same level of performance. While the deviations from a linear response may, in principle, be corrected, this kind of procedure is not possible in regions with no positional sensitivity and is also limited by the intrinsic resolution of the detector. Similar performance trends were observed for Micromegas and µRWELL amplification options, making zigzags widely applicable as well. In summary, by adopting an interleaved anode geometry early in the design stage of the readout presents a relatively straight forward approach for optimizing major performance considerations for specific detector applications, while maintaining a reasonable channel count. Alternatively, adding charge spreading mechanisms, like a DLC layer to a pre-existing pad layout increases the complexity of the system and may be prone to non-uniformities.

## V. Acknowledgements

We would like to thank the staff at the FTBF, including Mandy Rominsky, Eugene Schmidt, and Todd Nebel for their support and expertise during the beam test campaign.